\documentclass[aps,prl,showpacs,twocolumn,superscriptaddress,floatfix,tightenlines]{revtex4}
\pdfoutput=1 
\usepackage{bbm}
\usepackage{mathrsfs}
\usepackage{epsfig,psfrag}
\usepackage{amsmath,amsfonts,amssymb}
\usepackage[usenames]{color}
\usepackage{bm}

\newcommand{\cE}{\mathcal E}

\newcommand{\tr}{\text{Tr}}    

\definecolor{BrickRed}{cmyk}{0,0.89,0.94,0.28}%%%PANTONE 1805
\definecolor{MidnightBlue}{cmyk}{0.98,0.13,0,0.43}%%%PANTONE 302
\definecolor{DarkGreen}{rgb}{0,0.7,0.1}

\begin{document}

\preprint{draft}

\title{Casimir manipulations: The orientation dependence of
fluctuation-induced forces}

\author{T.\ Emig}
\affiliation{Institut f\"ur Theoretische Physik, Universit\"at zu K\"oln,
Z\"ulpicher Strasse 77, 50937 K\"oln, Germany}
\affiliation{Laboratoire de Physique Th\'eorique et Mod\`eles
Statistiques, CNRS UMR 8626, Universit\'e Paris-Sud, 91405 Orsay,
France}

\author{N.\ Graham}
\affiliation{Department of Physics, Middlebury College,
Middlebury, VT  05753} 

\author{R.\ L.\ Jaffe}
\affiliation{
Center for Theoretical Physics, Laboratory for Nuclear Science, and
Department of Physics, Massachusetts Institute of
Technology, Cambridge, MA 02139, USA}

\author{M.\ Kardar}
\affiliation{
Department of Physics, Massachusetts Institute of
Technology, Cambridge, MA 02139, USA} 
\date{\today}

\begin{abstract}
  The Casimir interaction between two objects, or between an object
  and a plane, depends on their relative orientations.  We make these
  angular dependences explicit by considering prolate or oblate
  spheroids.  The variation with orientation is calculated exactly at
  asymptotically large distances for the electromagnetic field, and at
  arbitrary separations for a scalar field.  For a spheroid in front
  of a mirror, the leading term is orientation independent, and we
  find the optimal orientation from computations at higher order.
\end{abstract}

\pacs{42.25.Fx, 03.70.+k, 12.20.-m. KITP report number: NSF-KITP-08-137}

\maketitle

Casimir forces, predicted in 1948 \cite{Casimir+CPolder} and
probed in high precision experiments over the past
decade \cite{Lamoreaux,Mohideen,Bressi,ChanSci}, are particularly
important at micro-meter to nano-meter length scales.  In constructing
and manipulating devices at these length scales it is important to
have an accurate understanding of the material, shape and orientation
dependence of these forces.  Since these dependencies often originate
in the inherent many-body character of the force, they
cannot be reliably obtained by commonly used approximations.  In this Letter we
investigate shape and orientation dependence of the Casimir
force using a recently developed method~\cite{us} (see also~\cite{Kenneth:2006vr}) 
that makes it possible to compute the Casimir interaction for arbitrary 
compact objects based on their scattering matrices.

A closely related work is the classic paper of Balian \&
Duplantier~\cite{Balian},
which provides expressions for the large distance asymptotic force
between perfectly conducting objects of arbitrary shape and
orientation.  Using our method, we generalize these results
in a form suitable for extension to arbitrary dielectrics and distances.  
As tangible examples we then focus on ellipsoids, computing the orientation
dependent force between two spheroids, and between a spheroid and a plane.  
The latter geometry is particularly interesting, as there in no orientation 
dependence {\em at leading order} for any object (metal or dielectric) in front of a mirror. 
We obtain the preferred orientation of a spheroid from a computation at higher order,
in which T-matrices are evaluated in an expansion around a spherical shape.
To numerically study a wider range of eccentricities and separations,
we consider a scalar theory (in place of electromagnetism), for which
efficient computations can be performed using a basis of spheroidal harmonics.
For the case of Neumann boundary conditions, we find a transition in the
preferred orientation as the spheroid approaches the plane.

We start from a general expression for the Casimir energy 
as an integral over imaginary wave number~\cite{us}, of $\tr \, \ln ( 1 - {\mathbb N})$,
which after expansion in powers of ${\mathbb N}$ reads
\begin{equation}
\label{eq:energy-scatt-exp2}
\cE = - \frac{\hbar c}{2\pi} \int_0^\infty d\kappa \,\left[
  \tr\left(\mathbb{N} \right) + \frac{1}{2}\tr\left(\mathbb{N}^2 \right)
+\cdots\right]\, .
\end{equation}
Here $\mathbb{N}=
\mathbb{T}^{1}\mathbb{U}^{12}\mathbb{T}^{2}\mathbb{U}^{21}$, 
where $\mathbb{T}^{1}$ and $\mathbb{T}^{2}$ relate incoming and
scattered electromagnetic (EM) fields, while the ``translation
matrix'' $\mathbb{U}^{12}$ relates the incoming wave
at one object to the outgoing wave at the other.  
Naturally, evaluating energies (and forces) from these
formal expressions requires judicious choice of basis.

A natural basis for the field is (vector) spherical harmonic
multipoles centered at each object, and labeled by $(l,m, \lambda)$;
the additional index $\lambda\in\{M,E\}$ distinguishes between
transverse electric (TE) or magnetic (TM) modes.   Each factor
of $\mathbb{U}$ decays as $e^{-\kappa d}$, so we can obtain an
asymptotic series, valid at large separations $d$.  Since
$\mathbb{T}^{\lambda\lambda}_{lm,l'm'}\propto \kappa^{l+l'+1}$, and
$\mathbb{T}^{\lambda\sigma}_{lm,l'm'}\propto \kappa^{l+l'+2}$ for
$\lambda\neq\sigma$, the expansion is dominated by the lowest order
multipoles.  Because electromagnetism does not admit monopole
fluctuations, the leading asymptotic behavior comes from $p$-waves,
$l=l'=1$.  By examining the response of an object to a uniform
electric (magnetic) field, it is straightforward to relate elements of
the $3\times3$ matrix $\mathbb{T}^{EE}_{1m,1m'}$
($\mathbb{T}^{MM}_{1m,1m'}$) to elements of the standard electric
(magnetic) polarizability matrix $\mathbb{\alpha}$
($\mathbb{\beta}$).  In terms of the Cartesian components of the latter
(with the $\hat{z}$ axis pointing from one object to the other), the
first asymptotic contribution to the energy is
\begin{eqnarray}
 \label{eq:energy_aniso}
 \cE_1^{12} &=&  -\frac{\hbar c}{d^7} \frac{1}{8\pi} \bigg\{
13\left( \alpha^1_{xx}\alpha^2_{xx} + \alpha^1_{yy}\alpha^2_{yy}+2 \alpha^1_{xy}\alpha^2_{xy}\right) 
\\
&+& 20 \, \alpha^1_{zz}\alpha^2_{zz} -30 \left( \alpha^1_{xz}\alpha^2_{xz} 
+ \alpha^1_{yz}\alpha^2_{yz}\right) + \left(\mathbb{\alpha}\to\mathbb{\beta}\right)
\nonumber\\
&-& 7 \left( \alpha^1_{xx}\beta^2_{yy} +  \alpha^1_{yy}\beta^2_{xx} 
-2 \alpha^1_{xy}\beta^2_{xy} \right) +\left( 1\leftrightarrow 2\right)
\bigg\}\, , \nonumber
\end{eqnarray}
which generalizes the result of  Balian and
Duplantier~\cite{Balian} for perfectly conducting objects
at nonzero temperatures.  For the case of an ellipsoidal object with
static electric permittivity $\epsilon$ and magnetic permeability $\mu$,
the polarizability tensors are diagonal in a basis oriented to its
principal axes, with elements (for $i\in\{1,2,3\}$)
\begin{equation}
\label{eq:pol-tensor-diag}
\alpha_{ii}^0 = \frac{V}{4\pi} \frac{\epsilon-1}{1+(\epsilon-1)n_i}\, ,\,
\beta_{ii}^0 = \frac{V}{4\pi} \frac{\mu-1}{1+(\mu-1)n_i}\,,
\end{equation}
where $V=4\pi r_1 r_2 r_3/3$ is the ellipsoid's volume and
the so-called depolarizing factors are given by
\begin{equation}
\label{eq:n-factors}
n_i = \frac{r_1 r_2 r_3}{2}\! \int_0^\infty \!\!\!
\frac{ds}{(s+r_i^2)\sqrt{(s+r_1^2)(s+r_2^2)(s+r_3^2)}} \,,
\end{equation}
in terms of the semi-axis dimensions $r_i$.
We can gain further insights into these results by focusing on the
case of spheroids, for which $r_1=r_2=R$ and $r_3 = L/2$.
Then the depolarizing factors can be expressed in terms of elementary
functions,
\begin{equation}
n_1=n_2=\frac{1-n_3}{2}, \, n_3 = \frac{1-e^2}{2e^3} \left(\log
\frac{1+e}{1-e} - 2 e \right),
\label{eq:depolarizing}
\end{equation}
where the eccentricity $e = \sqrt{1 - \frac{4R^2}{L^2}}$ is real for a
prolate spheroid ($L > 2R$) and imaginary for an
oblate spheroid ($L < 2R$).  The polarizability
tensors for an arbitrary orientation are then obtained as
$\mathbb{\alpha}={\cal R}^{-1}\mathbb{\alpha}^0{\cal R}$, where ${\cal
R}$ is the matrix that rotates the principal axis of the spheroid
to the Cartesian basis, i.e.  ${\cal R}(1,2,3)\to(x,y,z)$.
The result provides the leading Casimir energy for spheroids of
arbitrary size, orientation, and material.
Note that for rarefied media with $\epsilon\simeq 1$, $\mu\simeq 1$ the
polarizabilities are isotropic and proportional to the volume.
Hence, to leading order in $\epsilon-1$ the interaction is orientation
independent at asymptotically large separations, as we would expect,
since pairwise summation is valid for $\epsilon-1\ll 1$.
\begin{figure}[htbp]
\vspace*{-0.4cm}
\includegraphics[width=.98\linewidth]{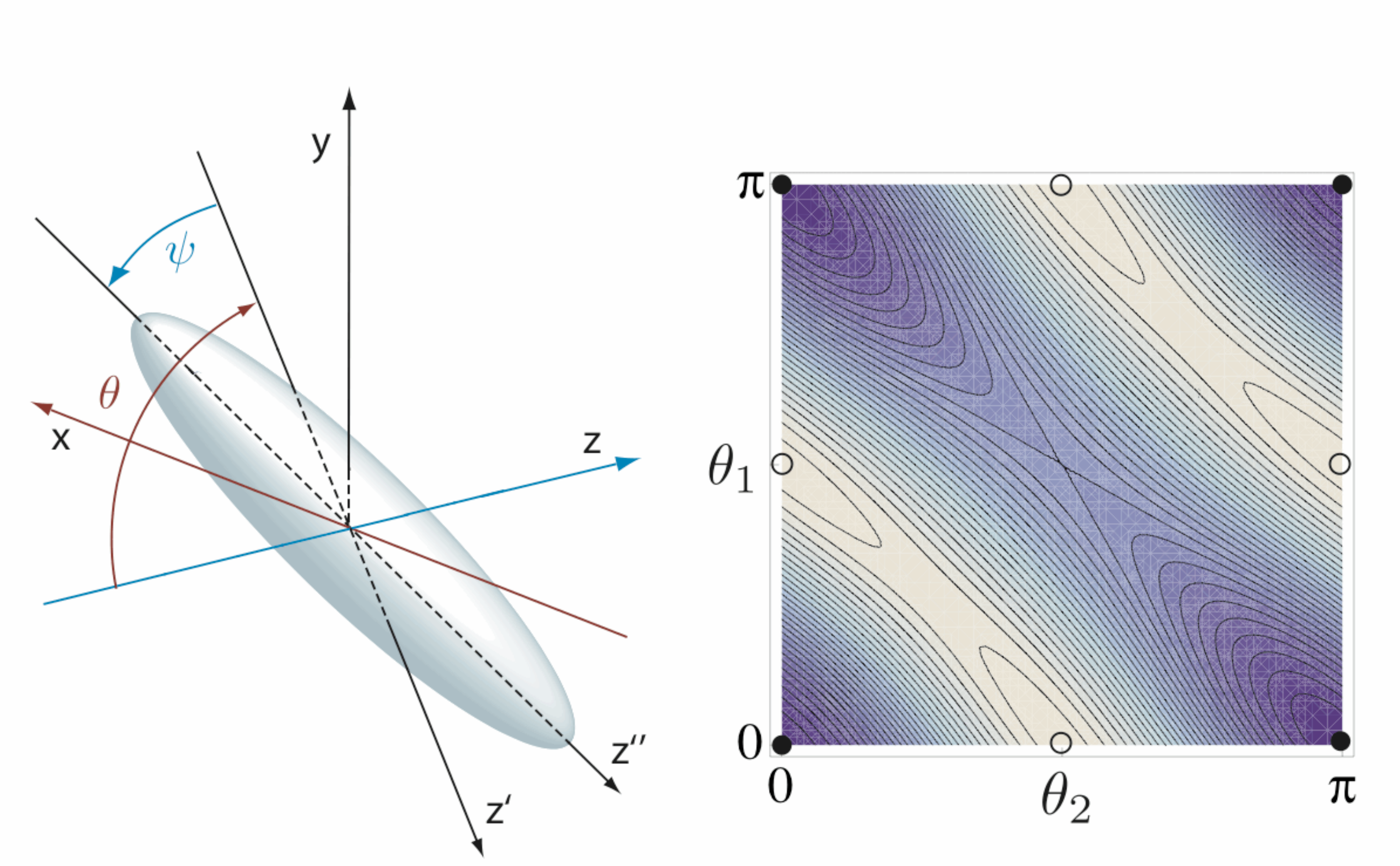}
\vspace*{-0.3cm}
\caption{\label{fig:cigars}Orientation of a prolate (cigar-shaped)
  spheroid: The symmetry axis (initially the $z$-axis) is rotated by
  $\theta$ about the $x$-axis and then by $\psi$ about the $z$-axis.
  For two such spheroids, the energy at
  large distances is give by Eq.~\eqref{eq:energy-cylidenr-general}.
 The latter is depicted at fixed distance $d$, and for
  $\psi_1=\psi_2$, by a contour plot as function
  of the angles $\theta_1$, $\theta_2$ for the $x$-axis rotations . 
  Minima (maxima) are marked by filled (open) dots.}
\end{figure}
\begin{figure}[htbp]
\includegraphics[width=\linewidth]{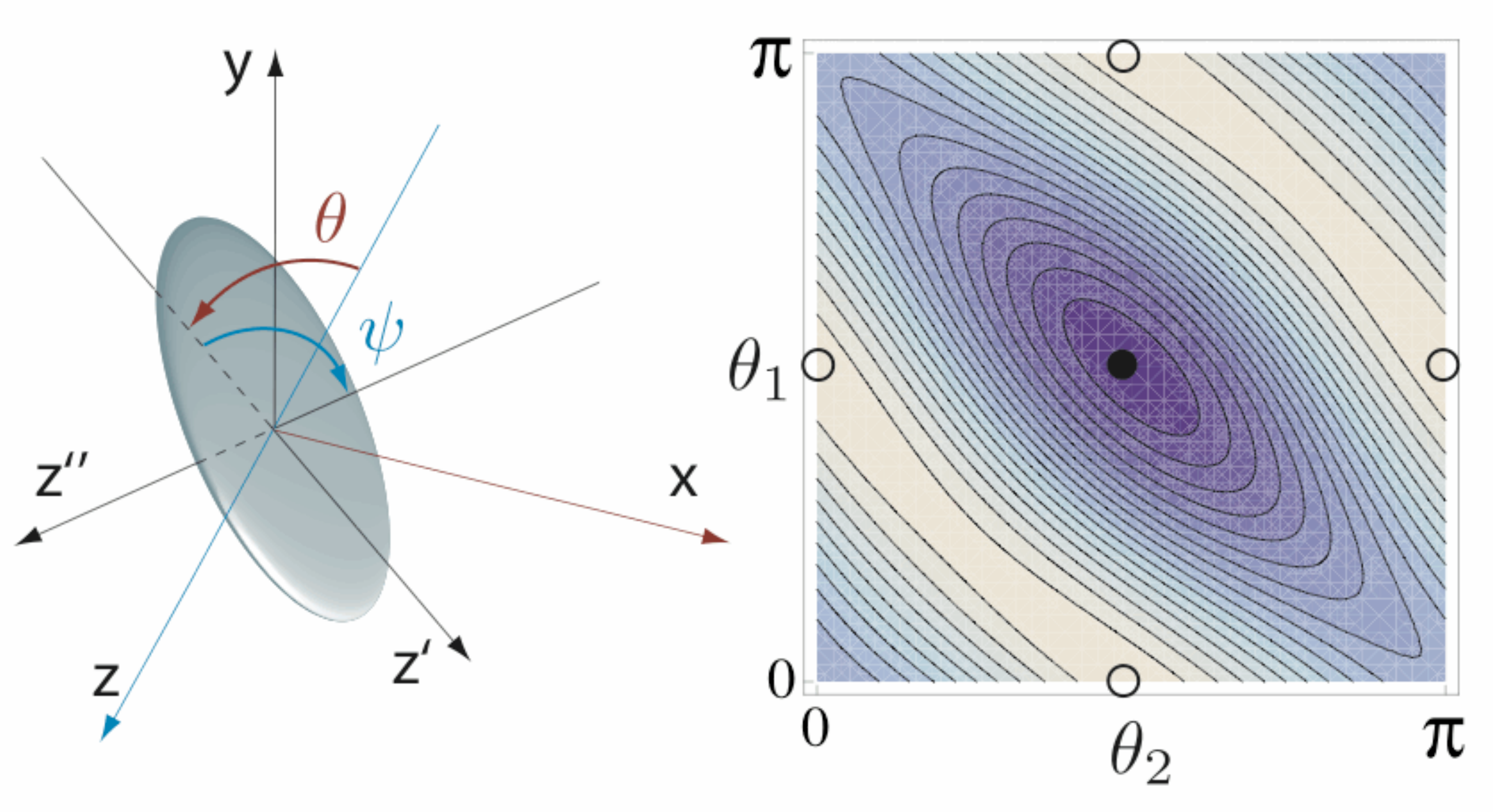}
\caption{\label{fig:pancakes} As in Fig.~\ref{fig:cigars} for oblate
  (pancake-shaped) spheroids, with a contour plot of energy at large
  separations.}
\end{figure}

As a special case, we present the explicit formula for 
identical perfectly conducting prolate spheroids with the length $L
\gg R$.  To implement the perfect conductor limit, we take $\epsilon
\to \infty$.  In this limit, the full T-matrix is independent of
$\mu$.  Because we are doing an expansion around zero frequency,
however, we must set $\mu=0$ to suppress the magnetic field that is
still allowed at zero frequency.  The orientation of each ``needle''
relative to the line joining them (the initial $z$-axis) is
parameterized by the two angles $(\theta,\psi)$, as depicted in
Fig.~\ref{fig:cigars}.  Then
\begin{eqnarray}
\label{eq:energy-cylidenr-general}
{\cal E}_1^{12}(\theta_1,\theta_2,\psi) &=& -\frac{\hbar c}{d^7} \bigg\{
\frac{5L^6}{1152 \pi \left( \ln \frac{L}{R} - 1\right)^2}
\times \\
&& \hspace*{-0.5cm}
\bigg[\cos^2\theta_1 \cos^2\theta_2
+ \frac{13}{20}\cos^2\psi \sin^2 \theta_1\sin^2\theta_2
\nonumber \\
&-& \frac{3}{8} \cos\psi \sin 2\theta_1 \sin 2\theta_2\bigg]
+{\cal O}\bigg(\frac{L^4R^2}{\ln\frac{L}{R}}\bigg)\bigg\}\, ,
\nonumber 
\end{eqnarray}
where $\psi\equiv\psi_1-\psi_2$.  

We note the following features of this result:
\par\noindent$\bullet$
The energy is minimized for $\theta_1=\theta_2=0$, i.e., for two
needles aligned parallel to their separation vector, as we
expect since this case is dominated by the fluctuations of
electric dipoles along the symmetry axis.
\par\noindent$\bullet$
At almost all orientations the energy scales as $L^6$, and vanishes
logarithmically slowly as $R\to 0$.
\par\noindent$\bullet$
The  dependence vanishes when one needle is orthogonal to
$\hat{z}$ (i.e. $\theta_1=\pi/2$), while the other is either parallel
to $\hat{z}$ ($\theta_2=0$) or has an arbitrary $\theta_2$ but differs
by an angle $\pi/2$ in its rotation about the $z$-axis
(i.e. $\psi_1-\psi_2=\pi/2$).  In these cases the energy comes
from the next order term in
Eq.~(\ref{eq:energy-cylidenr-general}), and takes the form
\begin{equation}
  \label{eq:crossed-cigars-finite-theta}
  {\cal E}_1^{12}\left(\frac{\pi}{2},\theta_2,\frac{\pi}{2}\right) = 
-\frac{\hbar c}{1152 \pi \, d^7} \frac{L^4R^2}{\ln\frac{L}{R} - 1} 
\left( 73+7\cos 2\theta_2
  \right) .
\end{equation}
\par\noindent$\bullet$
From Eq.~(\ref{eq:crossed-cigars-finite-theta}) we see that the
least favorable configuration ($\theta_2=\pi/2$) corresponds to two
needles orthogonal to each other and to the line joining them.

For perfectly conducting oblate spheroids with
$R\gg L/2$, the orientation of each ``pancake'' is again
described by a pair of angles $(\theta,\psi)$, as depicted in Fig.~\ref{fig:pancakes}.
A contour plot of the leading angular dependence is also presented in the figure;
its explicit formula will not be reproduced here, but we note the following features:
\par\noindent$\bullet$
The leading dependence is proportional to $R^6$, and does not disappear for any
choice of orientations.  Furthermore, this dependence remains even as
the thickness of the pancake is taken to zero ($L\to 0$). This is
very different from the case of the needles, where the interaction
energy vanishes with thickness as $\ln^{-1}(L/R)$. We attribute this
effect to dimensionality: The limiting linear needle is only
marginally visible to the EM field, while the limiting
two dimensional plane remains as an opaque obstacle.  
The lack of $L$ dependence is due to the assumed perfectly metallic  screening. 
If the dielectric function remains finite in the static
limit, a conventional scaling with volumes of the objects is expected.
\par\noindent$\bullet$
The configuration of minimal energy corresponds to two pancakes lying
on the same plane ($\theta_1=\theta_2=\pi/2$, $\psi=0$) and has energy
$-\hbar c \, (173/18\pi^3) R^6/d^7$.  This is due to the
electric dipole fluctuations in the planes.
\par\noindent$\bullet$
When the two pancakes are stacked on top of each other, the energy is
increased to $-\hbar c  \,(62/9\pi^3)  R^6/d^7$.
\par\noindent$\bullet$
The least favorable configuration is when the pancakes lie
in perpendicular planes, i.e., $\theta_1=\pi/2$, $\theta_2=0$, with an energy
$-\hbar c\, (11/3\pi^3) R^6/d^7 $.

For  an object interacting with a perfectly
reflecting mirror, we can use its image to construct
$\mathbb{N}= -\tilde {\mathbb T}^{1}\tilde
{\mathbb U}^{R,11}$, where $R$ stands for the reflected object~\cite{Emig:2008a}.  
Alternatively, we can express the scattering
matrix for the mirror in a basis of plane waves, and use a
translation matrix to convert between this basis and the
basis for scattering of the compact object.  At leading order we have
\begin{equation}
\label{eq:energy_aniso_wall}
\cE_1^{1m} = -\frac{\hbar c}{d^4} \frac{1}{8\pi} \tr (\alpha-\beta ) 
+{\cal O}(d^{-5})\, ,
\end{equation}
which is clearly independent of orientation. A similar expression
for a mirror and an atom is
present in the classic work of Casimir and Polder~\cite{Casimir+CPolder},
and implicit in Ref.~\cite{Balian} for perfect conductors.
(The orientation independence of energy is special to zero
temperature; at nonzero temperatures the general expression
in Ref.~\cite{Balian} does contain angular dependence.)
Orientation dependence in this system thus comes from higher
multipoles.  The next order also vanishes, so the leading term
is the $l=3$ contribution to the term linear in
${\mathbb N}$ in Eq.~\eqref{eq:energy-scatt-exp2},
in which case the scattering matrix is not known
analytically.

Instead, we obtained the preferred orientation by considering a distorted
sphere in which the radius $R$ is deformed to $R+\delta f(\vartheta,\varphi)$.
The function $f$ can be expanded into spherical harmonics $Y_{lm}(\vartheta,\varphi)$,
and spheroidal symmetry can be mimicked by choosing $f=Y_{20}(\vartheta,\varphi)$.
By performing a perturbative expansion of spherical T-matrices in $\delta$, the
leading orientation dependent part of the energy is obtained as
\begin{equation}
\cE_f = - \hbar c \frac{1607}{640 \sqrt{5} \pi^{3/2}} \frac{\delta R^4}{d^6} \cos(2\theta)   \,. 
\end{equation}
A prolate spheroid ($\delta>0$) thus minimizes its energy by pointing towards the mirror,
while an oblate spheroid ($\delta<0$) prefers to lie in a plane  perpendicular to the mirror.
(We assume that the perturbative results are not changed for large distortions.)
These configurations are also preferred at small distances $d$, since (at fixed
distance to the center) the object reorients to minimize the closest separation.

\begin{figure}[htbp]
\includegraphics[width=.9\linewidth]{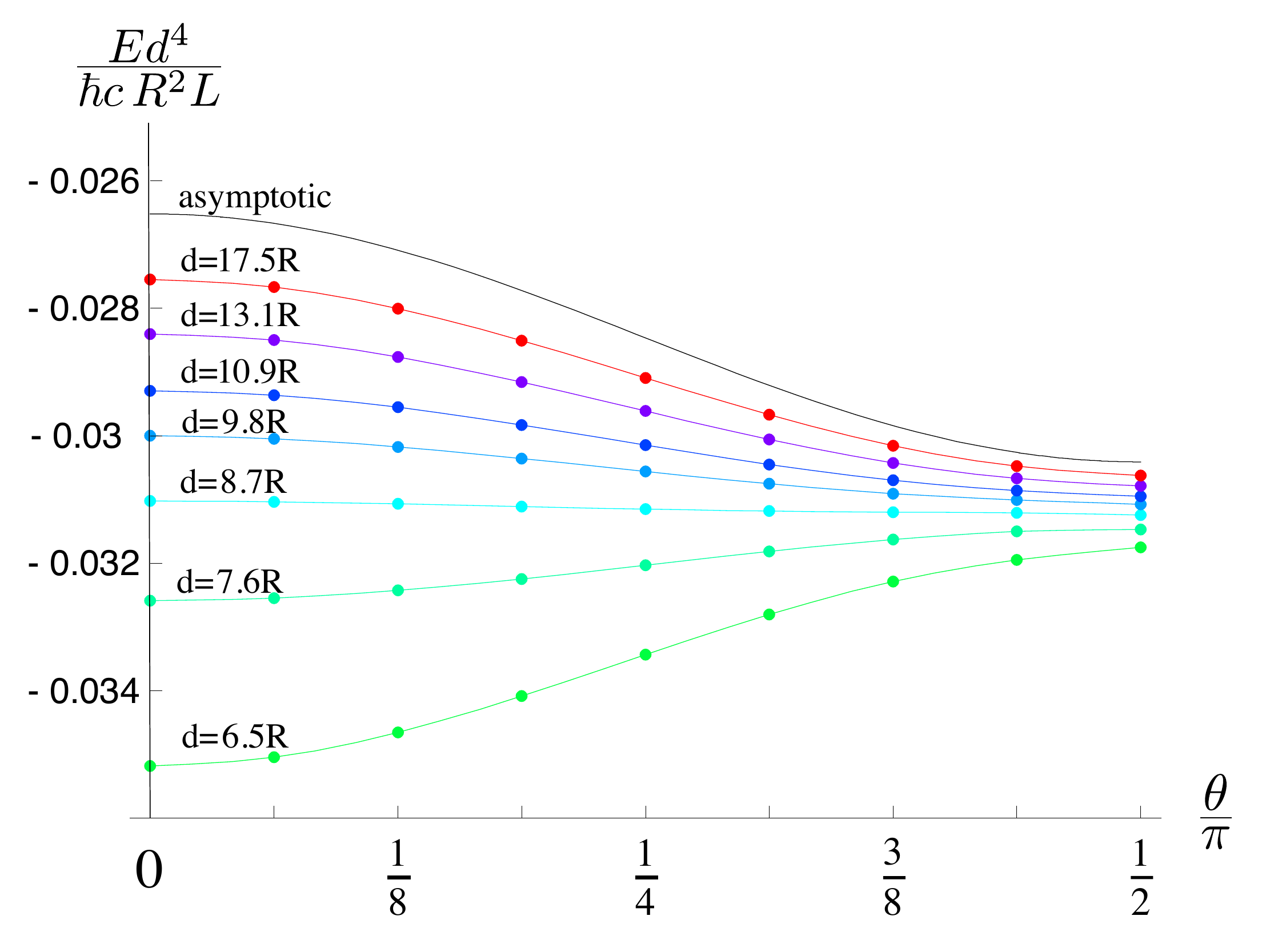}
\caption{\label{fig:spheroids} The orientation dependence of the
  Casimir energy for a prolate Neumann spheroid with $L/R=4.8$
  opposite a Neumann plane, as a function of the angle $\theta$
  between the normal to the plane and the spheroid axis, for various
  values of $d$, the distance from the spheroid center to the plane.
  The asymptotic result of Eq.~(\ref{eq:asymptotic}) is depicted by
  the top line.  For $d \lesssim 8.5 R$, the spheroid prefers to point
  toward the plane, while at larger distances it spontaneously breaks
  the rotational symmetry and aligns parallel to the plane.  The
  larger the value of $\xi_0=1/e$, the larger the distance at which
  this transition occurs.  }
\end{figure}

Interestingly, the above conclusion is not generally true, and as discussed
below, we find that there can be a transition in preferred orientation as
a function of $d$ in the simpler case of a scalar field.  Then
we can obtain the T-matrices and translation matrices as 
straightforward generalizations of the spherical case, in which
spherical harmonics and Bessel functions are replaced by
spheroidal harmonics and spheroidal radial functions.  
Prolate spheroidal coordinates~\cite{Flammer} are defined by
taking $\phi$ as the usual azimuthal angle and 
\begin{eqnarray}
\sqrt{x^2 + y^2} &=& a\sqrt{(\xi^2-1) (1-\eta^2)} ,\quad
z = a\eta\xi  \,,
\end{eqnarray}
where the interfocal separation is $2a$.  The surface of the spheroid
is defined by a constant spheroidal radius $\xi=\xi_0=1/e$, with
$L=2\xi_0 a$ and $R=a\sqrt{\xi_0^2 - 1}$.
Oblate coordinates are obtained by analytic
continuation of the radial coordinate with $\xi \to -i\xi$ and $k\to ik$. 
We use the conventions of Ref.~\cite{Graham+Olum} in which the spheroidal functions
are normalized identically to their spherical analogs.

The T-matrix is diagonal, with elements which in terms of radial
spheroidal functions of the first and third kinds are $T^S_{l m}(i
\kappa) = -\frac{R^{m(1)}_l{}'(i \kappa a;\xi_0)} {R^{m(3)}_l{}'(i
  \kappa a;\xi_0)}$, where the prime indicates derivative with respect
to the second argument.  The translation matrix converts between the
spheroidal basis, indexed by $l$ and $m$, and the plane wave basis,
indexed by the momentum parallel to the plane $\bm{q}_\perp$.  The
calculation is performed on the imaginary axis; with $q_\perp$ real,
and $q_\parallel \equiv i \sqrt{\kappa^2 + q_\perp^2}$, which is pure
imaginary.  The translation matrix element
$U_{lm}^{\bm{q}_\perp}(i\kappa)$ at separation $L$ is
$U_{lm}^{q_\perp, \phi_q} = i^l (-1)^m \frac{e^{i q_\parallel
    L}}{q_\parallel} {\cal Y}^{-m}_l \left(i\kappa a, \eta,
  \phi\right)$, where ${\cal Y}^{m}_l$ is a spheroidal harmonic and
$\eta = \frac{q_\parallel \cos \theta - q_\perp \cos \phi_q \sin
  \theta}{i\kappa}$.  Here, $\theta$ is the angle between the spheroid
axis and the normal to the plane and $\tan \phi = \frac{q_\perp \sin
  \phi_q} {q_\parallel \sin \theta + q_\perp \cos \theta \cos
  \phi_q}$, where $\phi_q$ is the polar angle of $\bm{q}_\perp$ around
the perpendicular to the plane.  To implement the matrix
multiplication needed to compute $\mathbb{N}$, we integrate over
$\bm{q}_\perp$.  Normalizing the integration measure introduces a
factor of $2 q_\parallel/(i\kappa)$.

The numerical results can be compared to the large distance 
expansion for an object with Neumann boundary conditions that is
placed opposite to a plane (the latter with Dirichlet or Neumann boundary
conditions).  The general form of the energy in this case is
\begin{equation}
  \label{eq:energy_nbc_plate}
   \cE = \pm \frac{\hbar c}{d^4} \left[ \frac{1}{64\pi^2} V 
- \frac{1}{16\pi} (\beta_{xx}+\beta_{yy}+3\beta_{zz})\right] \, ,
\end{equation}
for a Dirichlet/Neumann $(+/-)$  plane, where $V$ is the volume,
and $\beta$ is like the static magnetic polarizability tensor in Eq.~(\ref{eq:pol-tensor-diag}) 
for $\mu=0$. For a spheroid this results in
\begin{equation}
\cE = \pm \frac{\hbar c}{d^4} \frac{R^2 L}{96 \pi}
\frac{9 - 4n_3 - n_3^2 + (3 n_3 -1) \cos 2\theta }{1-n_3^2} \,,
\label{eq:asymptotic}
\end{equation}
where $n_3$ is given by Eq.~(\ref{eq:depolarizing}).
In contrast to the EM energy of
Eq.~\eqref{eq:energy_aniso_wall}, the leading term of
Eq.~(\ref{eq:asymptotic}) is orientation dependent.  For a prolate
spheroid with $R \ll L$ we have
\begin{equation}
\label{eq:energy_nbc_prolate_plate}
\cE = \pm \frac{\hbar c}{d^4} \frac{R^2 L}{96\pi} 
\left[ 9 -\cos(2\theta)  +{\cal O}((R/L)^2)\right] \, ,
\end{equation}
and for an oblate spheroid with $R\gg L$, we have
\begin{equation}
\label{eq:energy_nbc_oblate_plate}
\cE = \pm \frac{\hbar c}{d^4} \frac{R^3}{24\pi^2} 
\left[ 2+\cos(2\theta)
+{\cal O}((L/R))\right] \, .
\end{equation}
To leading order in $R/L$ and at fixed separation $d$, the energy for
a Neumann plate is minimized at $\theta=\pi/2$ for a prolate spheroid
and at $\theta=0$ for an oblate spheroid.  This means that both a
needle and a pancake prefer to be parallel to the plate;
in the former case because of $\beta_{xx}$ and
$\beta_{yy}$ and in the latter case because of
$\beta_{zz}$.  For a Dirichlet plate, the energy is minimal if the
needle or the pancake are perpendicular to the plate 
as would have been predicted by pairwise summation.

In Fig.~\ref{fig:spheroids} we present a sample of the full numerical
calculation, using an optimized version~\cite{home} of the spheroidal 
harmonic package of Falloon~\cite{Falloon:2002}.
We truncate the infinite sum at $l=2$, though we
have verified for individual cases that summing through $l=4$ does not
significantly change these results.  At close separations, the prolate
spheroid prefers to point toward the plane, as we expect from
the proximity force approximation.  At large separations, we have
shown that the configuration parallel to the plane has the lowest
energy.  We thus observe a spontaneous breaking of the symmetry around
the perpendicular to the plane; the separation at which this transition
occurs varies with the spheroid's eccentricity.

These examples provide a glimpse into the myriad
effects of shape and orientation on the Casimir effect. One can
imagine that with better precision of experiment and theory such
forces can be employed to manipulate small objects at the
micro- and nano-meter scales.

We acknowledge discussions with S.~J.~Rahi.
This research was supported by the DFG through 
grant EM70/3 (TE), and NSF grants DMR-08-03315 (MK),
PHY-0555338 (NG), a Cottrell College Science Award from Research
Corporation (NG), and the U.~S.~Department of Energy under
cooperative research agreement \#DF-FC02-94ER40818 (RLJ).
Part of this work was carried out at the Kavli Institute for
Theoretical Physics, with support from NSF Grant No. PHY05-51164.

\end{document}